\documentclass{osa-article}

\journal{oe}
\articletype{Research Article}
\usepackage[T1]{fontenc}
\usepackage{ae,aecompl}
\usepackage[]{graphicx}
\usepackage{amsmath}
\usepackage{lmodern}
\usepackage{url}
\usepackage{breqn}
\usepackage{siunitx}
\usepackage{color, soul}
\usepackage{subcaption}
\usepackage[frozencache]{minted}
\usepackage{todonotes}

\newsavebox{\mintedbox}
\newenvironment{mintedsubcaptionbox}[2]
 {%
  \VerbatimEnvironment
  \def\mscbcaption{#1}%
  \RecustomVerbatimEnvironment{Verbatim}{BVerbatim}{}%
  \begin{lrbox}{\mintedbox}%
  \begin{minted} {#2}%
 }
 {%
  \end{minted}%
  \end{lrbox}%
  \subcaptionbox{\mscbcaption}[1.\textwidth]{\usebox{\mintedbox}}\\%
 }

\usepackage{acronym}
\acrodef{AO}[AO]{Adaptive Optics}
\acrodef{DM}[DM]{Deformable Mirror}
\acrodef{FFT}[FFT]{Fast Fourier Transform}
\acrodef{LGS}[LGS]{Laser Guide Star}
\acrodef{PSF}[PSF]{Point Spread Function}
\acrodef{PYPI}[PyPI]{Python Package Index}
\acrodef{RTC}[RTC]{Real-Time Control}
\acrodef{SLODAR}[SLODAR]{Slope Detection And Ranging}
\acrodef{STFC}[STFC]{Sciences and Technology Funding Council}
\acrodef{WFS}[WFS]{Wavefront Sensor}

\begin{document}

\title{AOtools - a Python package for adaptive optics modelling and analysis}

\author{M. J. Townson,\authormark{1,*} O. J. D. Farley,\authormark{1} G. Orban de Xivry,\authormark{2} J. Osborn,\authormark{1} and A. P. Reeves\authormark{3}}

\address{\authormark{1}Department of Physics, Durham University, South Road, DH1 3LE, UK\\
\authormark{2}Space sciences, Technologies \& Astrophysics Research (STAR) Institute, University of Li\`ege, Li\`ege, Belgium\\
\authormark{3}Institute   of   Communications   and   Navigation, German Aerospace Center (DLR), M\"{u}nchener Stra{\ss}e 20, 82234 We{\ss}ling, Germany}

\email{\authormark{*}matthew.townson@durham.ac.uk} %% email address is required



\begin{abstract}
AOtools is a Python package which is open-source and aimed at providing tools for adaptive optics users and researchers.
We present version 1.0 which contains tools for adaptive optics processing, including analysing data in the pupil plane, images and point spread functions in the focal plane, wavefront sensors, modelling of atmospheric turbulence, physical optical propagation of wavefronts, and conversion between frequently used adaptive optics and astronomical units.
The main drivers behind AOtools is that it should be easy to install and use.
To achieve this the project features extensive documentation, automated unit testing and is registered on the Python Package Index.
AOtools is under continuous active development to expand the features available and we encourage everyone involved in adaptive optics to become involved and contribute to the project.
\end{abstract}

%%%%%%%%%%%%%%%%%%%%%%%%%  body  %%%%%%%%%%%%%%%%%%%%%%%%%
% \section{Abstract}
% The abstract should be limited to approximately 100 words.
% If the work of another author is cited in the abstract, that citation should be written out without a number, (e.g., journal, volume, first page, and year in square brackets [Opt. Express {\bfseries 22}, 1234 (2014)]).

% \section{Submission}
% After proofreading the manuscript, compress your .tex manuscript file and all figures (which should be in EPS or PDF format) in a ZIP, TAR or TAR-GZIP package.

\section{Introduction}
\label{sect:intro}
The Python programming language (\url{http://www.python.org}) has become a powerful tool for scientific research \cite{Lima2006,Millman2011}.
% Currently the Python programming language  is a fast growing programming language in Astronomy and related fields \cite{Greenfield2011}.
Currently Python uptake in astronomy and related fields is growing rapidly \cite{Greenfield2011}.

\ac{AO} has reached a level of maturity where many algorithms and techniques are considered ``standard practice'', with cookbooks for implementing many standard procedures existing to help users implement functions, such as  \cite{Schmidt2010}.
There are freely available tools for \ac{AO} analysis on GPU (MOAO framework \cite{Abdelfattah2014}).
There are also a number of full end-to-end simulations which implement full \ac{AO} systems in software, such as the C/Python based DASP \cite{Basden2007}, Matlab\textsuperscript{\tiny\textregistered} based OOMAO \cite{Conan2014}, Yorick based YAO \cite{Rigaut2013b} and IDL based CAOS \cite{Carbillet2005}.

However, currently there is no freely available, open-source tool which implements a library for use in either simulation or \ac{AO} data analysis in Python.
We propose AOtools as a project and Python package which encompasses this functionality.
One example of using AOtools to create a full end-to-end simulation is the SOAPY project \cite{Reeves2016}. 

For the project to be successful it is essential that the AOtools package is reliable, easy to install, and easy to use.
This mandates the use of automation tools to ensure documentation is up-to-date and that the core functionality of the code is working as expected.
To this end infrastructure has been put in place to make the package ``pip'' installable, with minimal external dependencies.
Currently the only dependencies AOtools requires are: Numpy\cite{Vanderwalt2011}, Scipy\cite{Jones}, and Matplotlib\cite{Hunter2007}, all of which are standard tools of the scientific Python ecosystem.

In this paper we outline the major functionality of the software at the current release (v1.0), the structure of the project, the approach used for development of the library, and finally a road map for the future of the project.
It is envisaged that this project will be useful beyond the realm of astronomical \ac{AO}, including microscopy \cite{Bille1990, Marcos2017} and free space optical communications \cite{Dikmelik05}.

\section{Functionality}
In this section we present the core functionality of the AOtools package, covering functions applied to pupils in Sec.~\ref{sec:pupil}, simulation of atmospheric turbulence in Sec.~\ref{sec:turbulence}, processing \ac{AO} images and \ac{WFS} data in Sec.~\ref{sec:image} and \ref{sec:wfs} respectively, physical optical propagation kernels in Sec.~\ref{sec:propagation}, and conversions between astronomical units frequently used in \ac{AO} in Sec.~\ref{sec:conversions}.

To demonstrate the use of AOtools a snippet of code is presented along with plots which shows how to generate the data for the corresponding figure.
Working examples of all of the code used to generate the plots shown in this paper can also be found at \url{https://github.com/AOtools/aotools_tests}.
Finally, an example of the documentation provided by AOtools is shown in Fig.~1(c).

% Pupil Functions
\subsection{Pupil functions}
\label{sec:pupil}
\subsubsection{Circular masks}
\begin{figure}
\begin{mintedsubcaptionbox}{Code snippet.}{python}
>>> import aotools
>>> pupil_mask = aotools.circle(64, 128) - aotools.circle(16, 128)
\end{mintedsubcaptionbox}
\subcaptionbox{Output pupil mask.}[1.\textwidth]
        {\includegraphics[width=0.5\textwidth]{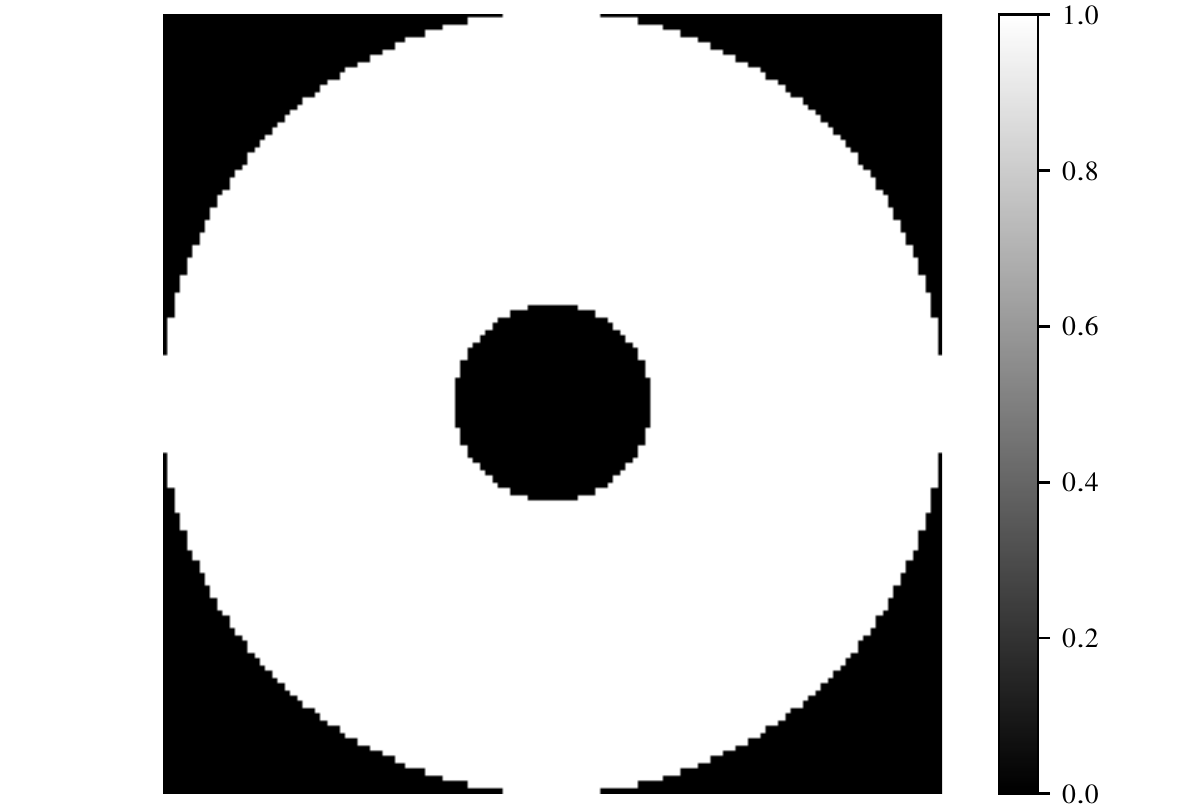}}
\subcaptionbox{Documentation example.}[1.\textwidth]
        {\includegraphics[width=0.7\textwidth]{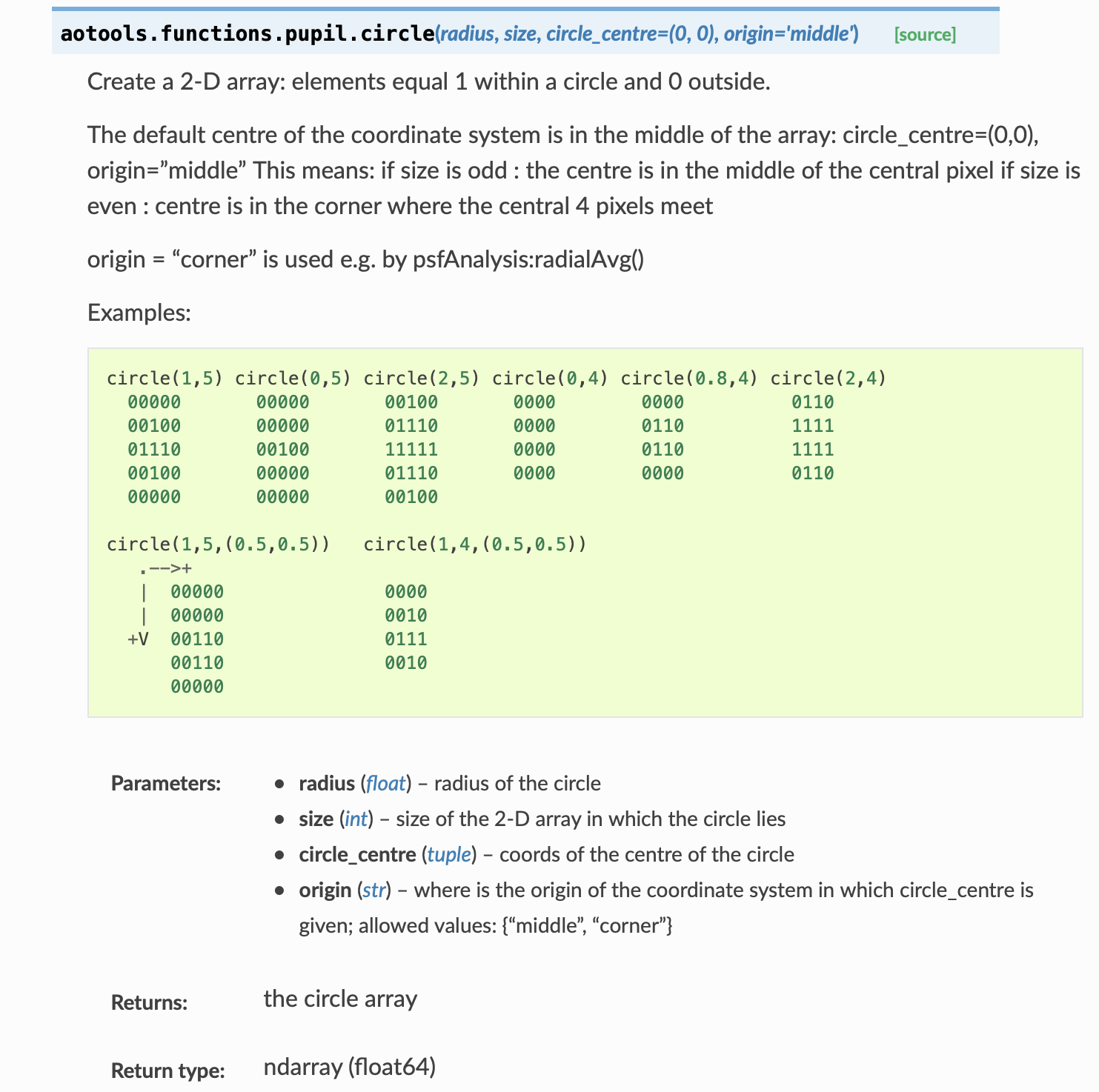}}
\caption{\label{fig:circle}
		 Example code snippet Fig.~1(a) for the creation of a telescope pupil mask using two circles to create a circular aperture with a cut-out section corresponding to the secondary mirror. The pupil mask created is shown in Fig.~1(b).
		 An example of the documentation provided by AOtools for this function is shown in Fig.~1(c).
		 }
\end{figure}

Generating circular functions in 2-D arrays is useful in generating pupil masks, as well as other applications.
Most lenses can be approximated with a circular pupil mask, and telescope pupil shapes can be approximated simply with one or two circular apertures, to create the obstruction of the secondary mirror.
A simple function exists in AOtools to create circular apertures, which can be combined to generate simple telescope pupil masks.
An example code snippet for the generation of a telescope pupil mask is shown in Fig.~\ref{fig:circle}(a), with the corresponding generated telescope pupil mask shown in Fig.~\ref{fig:circle}(b).

\subsubsection{Zernike modes}
\label{sec:zernike}
\begin{figure}
\begin{mintedsubcaptionbox}{Code snippet.}{python}
>>> import aotools
>>> number_of_modes = 5
>>> pupil_size = 128
>>> zernike_array = aotools.zernikeArray(number_of_modes, pupil_size)
\end{mintedsubcaptionbox}
\subcaptionbox{Output Zernike modes.}[1.\textwidth]
    {\includegraphics{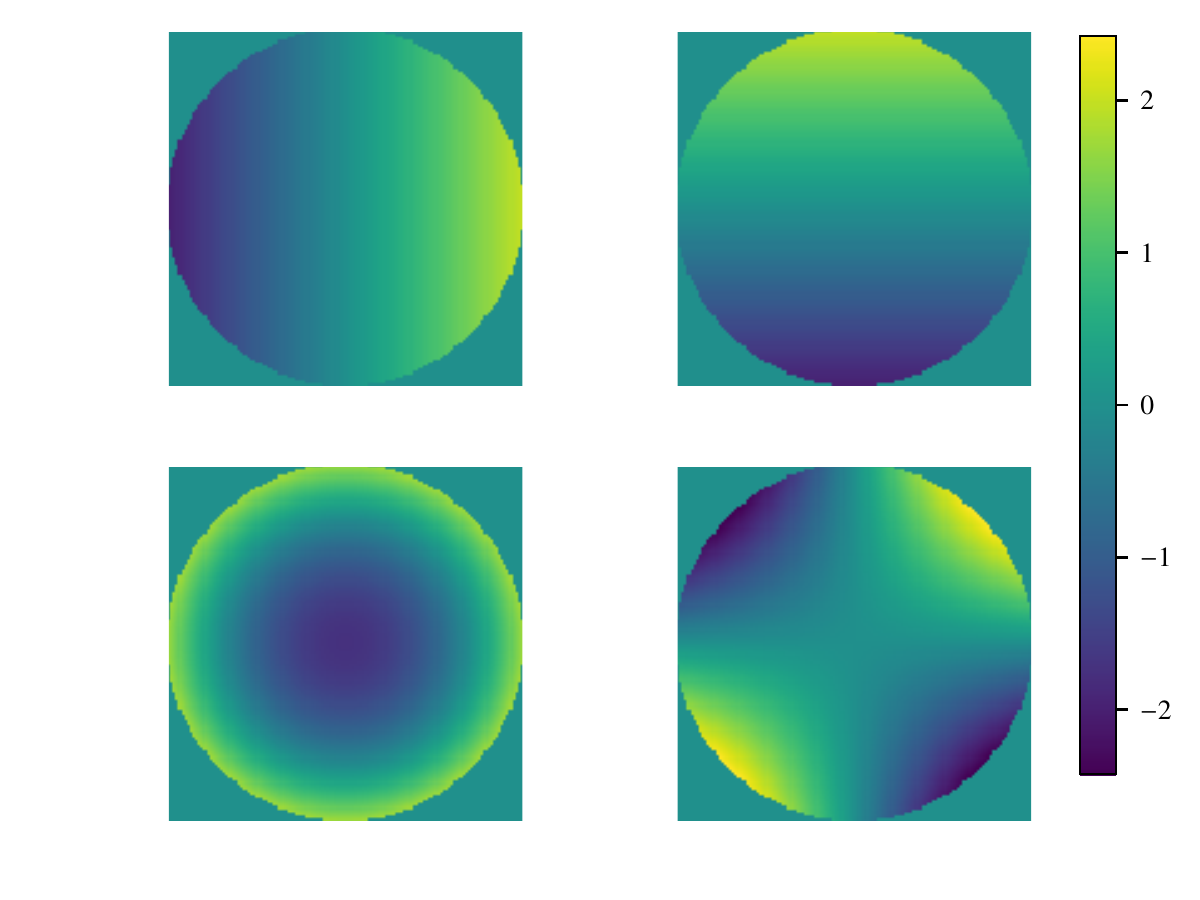}}
\caption{\label{fig:zernike}
    Fig.~2(a) shows the creation of the first five Zernike modes (piston, tip, tilt, defocus, and astigmatism) using the AOtools Zernike functions on circular apertures.
	This function creates all Zernikes up to the given J index.
	The output array of these modes, excluding piston, is shown in Fig.~2(b).
	}
\end{figure}

It is frequently useful to decompose wavefronts into Zernike modes.
This allows for analysis of aberrations and the performance of correction in \ac{AO} systems, for example atmospheric statistics are frequently decomposed into Zernike modes \cite{Noll1976}.
AOtools defaults to normalising individual modes using Noll, however, they can also be normalised by peak-to-valley and RMS strength if required.
An example snippet for creating an array of Zernike modes is given in Fig.~\ref{fig:zernike}(a), with a selection of the generated Zernike modes plotted in Fig.~\ref{fig:zernike}(b).

% Atmospheric Turbulence
\subsection{Atmospheric turbulence}
\label{sec:turbulence}
\begin{figure}
\begin{mintedsubcaptionbox}{Code snippet.}{python}
>>> from aotools.turbulence.infinitephasescreen import PhaseScreenKolmogorov
>>> # Set up parameters for creating phase screens
>>> nx_size = 128
>>> D = 8.
>>> pxl_scale = D/nx_size
>>> r0 = 0.164
>>> L0 = 100
>>> stencil_length_factor = 32
>>> # Create the initial phase screen
>>> phase_screen = PhaseScreenKolmogorov(nx_size, pxl_scale, r0, L0,
                                         stencil_length_factor)
>>> # Move phase screen along by a single row
>>> phase_screen.add_row()
\end{mintedsubcaptionbox}
\subcaptionbox{Output phase screen.}[1.\textwidth]
    {\includegraphics{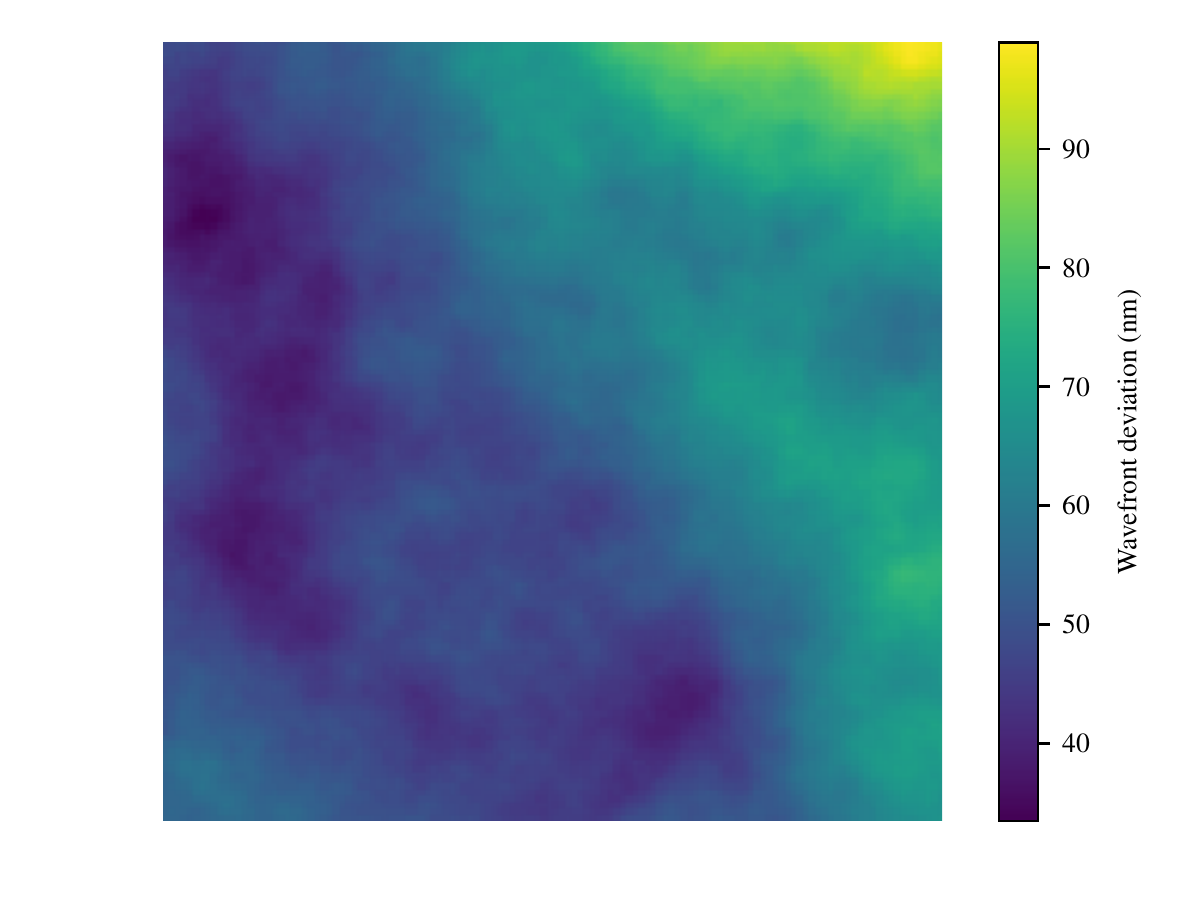}}
\caption{\label{fig:phase_screen}
                Fig.~3(a) shows an example code snippet with the specification and generation of an infinite phase screen using the Fried method \cite{Assemat2006, Fried2008}.
                The phase screen can be ``moved'' by using the \texttt{add_row} method which is shown at the end of the snippet.
                An example phase screen generated using this code is shown in Fig.~3(b).
		        The strength is defined in $\si{\nano\meter}$ such that the phase screen is wavelength independent.}
\end{figure}

\begin{figure}
	\centering
	\includegraphics{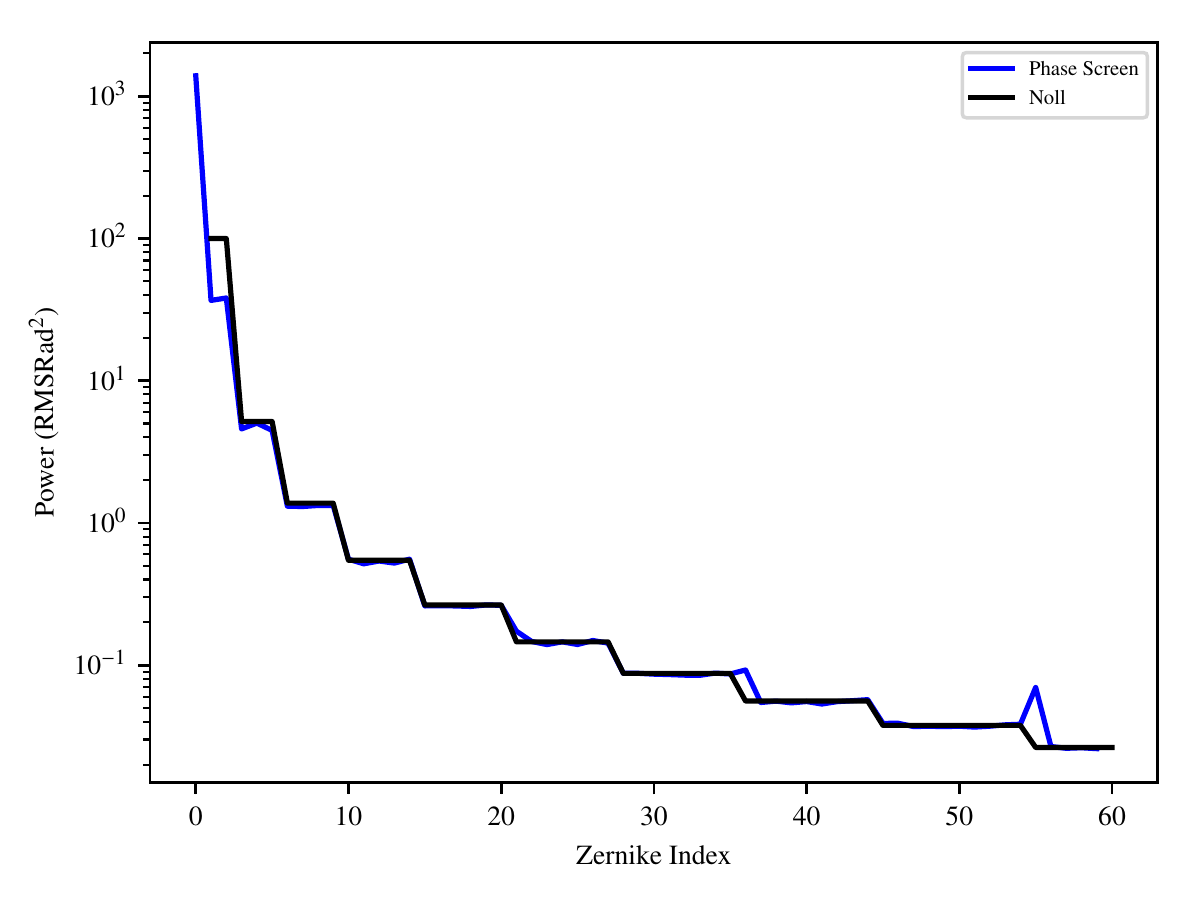}
	\caption{\label{fig:zernike_breakdown}
	The average Zernike breakdown of 1000 individual Fried infinite phase screens \cite{Assemat2006, Fried2008}, with the theoretical strengths expected from Noll \cite{Noll1976}.
	}
\end{figure}

% \subsubsection{Phase Screens}
AOtools features the functionality to create both finite Kolmogorov \cite{Kolmogorov1941, Klyatskin1972} and von K\'arm\'an turbulence phase screens.
These functions are based on the example implementations given in \cite{Schmidt2010}.
The library also includes methods to create infinite Kolmogorov and von K\'arm\'an phase screens using the methods described in \cite{Assemat2006}.
The phase screens created are wavelength independent as the amplitude scale is in $\si{\nano\meter}$.
An example code snippet using the Fried method for generating infinite phase screens \cite{Assemat2006, Fried2008} is shown in Fig.~\ref{fig:phase_screen}(a), with an image of the corresponding phase screen given in Fig.~\ref{fig:phase_screen}(b).
% An example phase screen generated using the Fried method for generating infinite phase screens \cite{Assemat2006, Fried2008} is shown in Fig.~\ref{fig:phase_screen} with the corresponding code used to generate it given in Fig.~\ref{code:phase_screen}.
The average Zernike breakdown of 1000 Fried infinite phase screens is shown in Fig.~\ref{fig:zernike_breakdown} alongside the theoretical amplitudes of each mode from \cite{Noll1976}.

% The phase screens have also been verified for scaling appropriately for  $r_{0}$ by measuring the total variance within the phase screen. \todo{Make plot showing $r_0$ scaling}

\subsection{Image processing}
\label{sec:image}

\subsubsection{Point source}
\begin{figure}
\begin{mintedsubcaptionbox}{Code snippet.}{python}
>>> import aotools
>>> import numpy
>>> # Create padding to get an oversampled psf at the end to make it look nice
>>> padded_pupil = numpy.zeros((1024, 1024))
>>> padded_pupil[:128, :128] = pupil_mask
>>> # Use AOtools to transform from the pupil to the focal plane
>>> psf = aotools.ft2(padded_pupil, delta=1./128.,)
>>> # Cut out the centre of the psf array for plotting
>>> psf = psf[512-64:512+64, 512-64:512+64]
>>> # Calculate the azimuthal average of the PSF
>>> azi_avg = aotools.azimuthal_average(numpy.abs(psf))
\end{mintedsubcaptionbox}
\subcaptionbox{Output \ac{PSF}.}[1.\textwidth]
    {\includegraphics{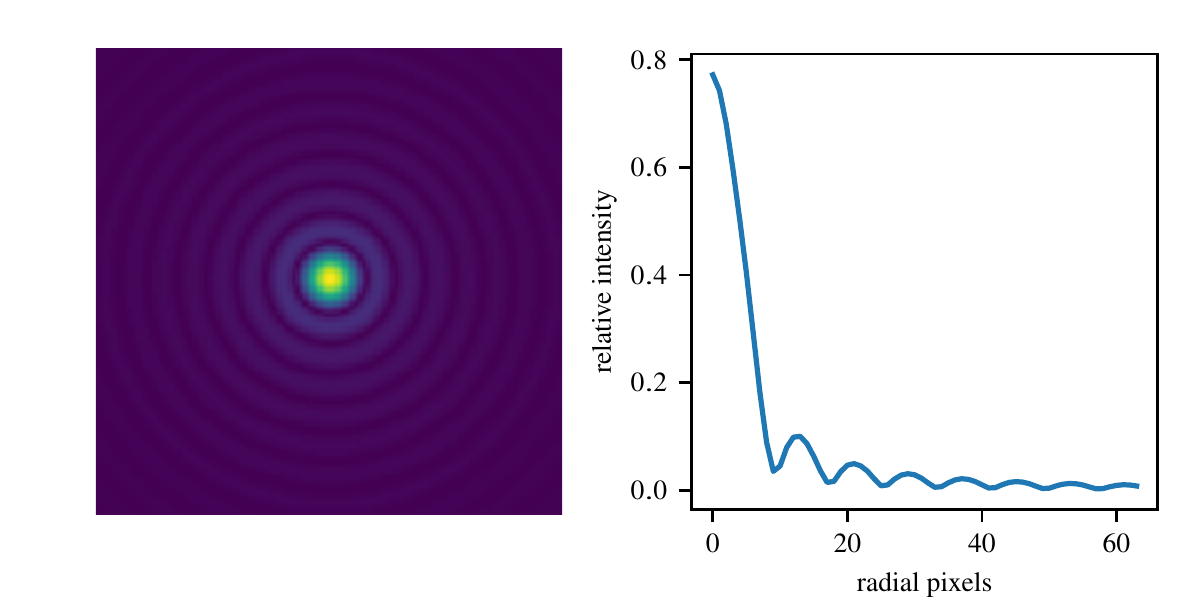}}
\caption{\label{fig:psf}
                 Example code snippet Fig.~5(a) to generate a \ac{PSF} from the pupil mask shown in Fig.~\ref{fig:circle}(b) and the calculation of the azimuthal average of the resulting \ac{PSF}.
                 The padding and slicing of the pupil mask is included to oversample the FFT, leading to a smoother \ac{PSF} for plotting.
                 The \ac{PSF} corresponding to the pupil shown in Fig.~\ref{fig:circle}(b) is shown alongside the corresponding azimuthal average of the \ac{PSF} calculated using AOtools in Fig.~5(b).}
\end{figure}

AOtools includes some functions for analysing \acp{PSF}.
This includes the full encircled energy as a function of radius, and the radius corresponding to a given fraction of encircled energy.
There is also the functionality to calculate the azimuthal average of an image.
However, currently this is restricted to the \ac{PSF} being in the centre of the input 2-D image.
An example of generating a \ac{PSF} and then using the azimuthal averaging function is given in Fig.~\ref{fig:psf}(a).
The \ac{PSF} generated from the pupil shown in Fig.~\ref{fig:circle}(b) and the azimuthal averaged intensity is shown in Fig.~\ref{fig:psf}(b).

AOtools also features a number of methods for centroiding images of point sources and elongated \acp{LGS}.
These include the centre-of-mass method, including parameters for thresholding within the centroider, and other methods used in \acp{WFS}, such as the brightest pixel centroider \cite{Basden2012}.
Example code snippet for centroiding an image is given in Fig.~\ref{code:cog}.
\begin{figure}
\begin{minted} {python}
>>> from aotools import centroiders
>>> cog_centroid = centroiders.centre_of_gravity(psf)
>>> bpx_centroid = centroiders.brightest_pixel(psf)
\end{minted}
\caption{\label{code:cog}
	     Centroiding a point source image using AOtools built-in centroiding functions on the \ac{PSF} shown in Fig.~\ref{fig:psf}(b).
	     These functions give centroids in the units of array indices (with [0,0] representing the corner of the image), so the centroid location corresponds to the index of the array.
	     }
\end{figure}

\subsubsection{Extended source}
\begin{figure}
\begin{mintedsubcaptionbox}{Code snippet.}{python}
>>> from aotools import centroiders
>>> from astropy.io import fits
>>> ext_image = fits.getdata("extended_image.fits")
>>> ref_image = fits.getdata("reference_image.fits")
>>> # Simple centroid for extended objects
>>> centroid = centroiders.correlation_centroid(ext_image,ref_image)
>>> # Manual correlation and centroiding of extened object
>>> corr_image = centroiders.cross_correlate(ext_image, ref_image)
>>> bpx_centroid = centroiders.brightest_pixel(corr_image)
\end{mintedsubcaptionbox}
\subcaptionbox{\ac{WFS} images and output cross-correlation image.}[1.\textwidth]
    {\includegraphics{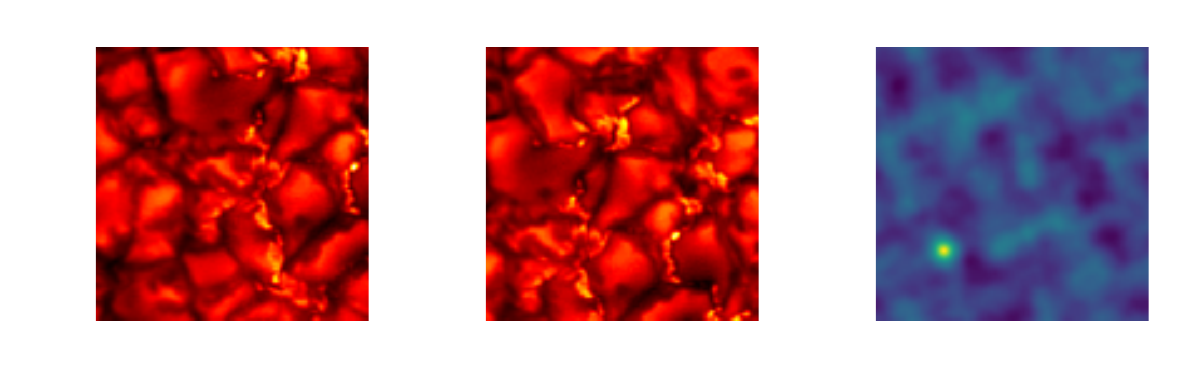}}

\caption{\label{code:extended_centroid}
         Example use of AOtools to centroid extended objects.
         AOtools can be used to centroid using a cross-correlation and a centre-of-mass using a single function call, or the correlation image can be generated in one step then any centroiding technique can be applied to the correlation image, as shown in Fig.~7(a).
         In Fig.~7(b) the left and middle images show the reference image and a shifted sub-aperture image of solar granulation respectively.
             The right most image shows the correlation image created from the cross-correlation of the left and centre images.
             This can then be centroided in order to measure the shift between the sub-aperture and the reference images.
         }
\end{figure}

AOtools also includes methods for centroiding extended objects using a cross-correlation technique.
The correlation images can be created using a built-in cross-correlation function, then the resulting correlation image centroided using any centroiding function \cite{Luhe1983}.
Or, if a simple centre-of-mass centroider will be applied to the correlation image, there is a \texttt{correlation_centroid} function which will perform both the cross-correlation and the centroiding to output the relative image shifts between images and a reference image.
Currently the only available method for cross-correlation is based on \acp{FFT}.
The use of AOtools to centroid extended objects is shown in Fig~\ref{code:extended_centroid}(a), with the example images, and correlation image generated shown in Fig~\ref{code:extended_centroid}(b).
There is also the functionality to measure the contrast of images of extended objects, with both the Michelson \cite{Michelson1927} and RMS contrast currently implemented.

\subsection{Wavefront sensors}
\label{sec:wfs}
\begin{figure}
\begin{mintedsubcaptionbox}{Code snippet.}{python}
>>> import aotools
>>> sub_apertures = 7
>>> fill_factor = 0.6
>>> sub_ap_width = 128./sub_apertures
>>> # Create active sub-aperture array
>>> sub_aps = aotools.wfs.findActiveSubaps(sub_apertures, pupil_mask,
                                           fill_factor)
\end{mintedsubcaptionbox}
\subcaptionbox{Sub-aperture grid overlaid on pupil image.}[1.\textwidth]
    {\includegraphics{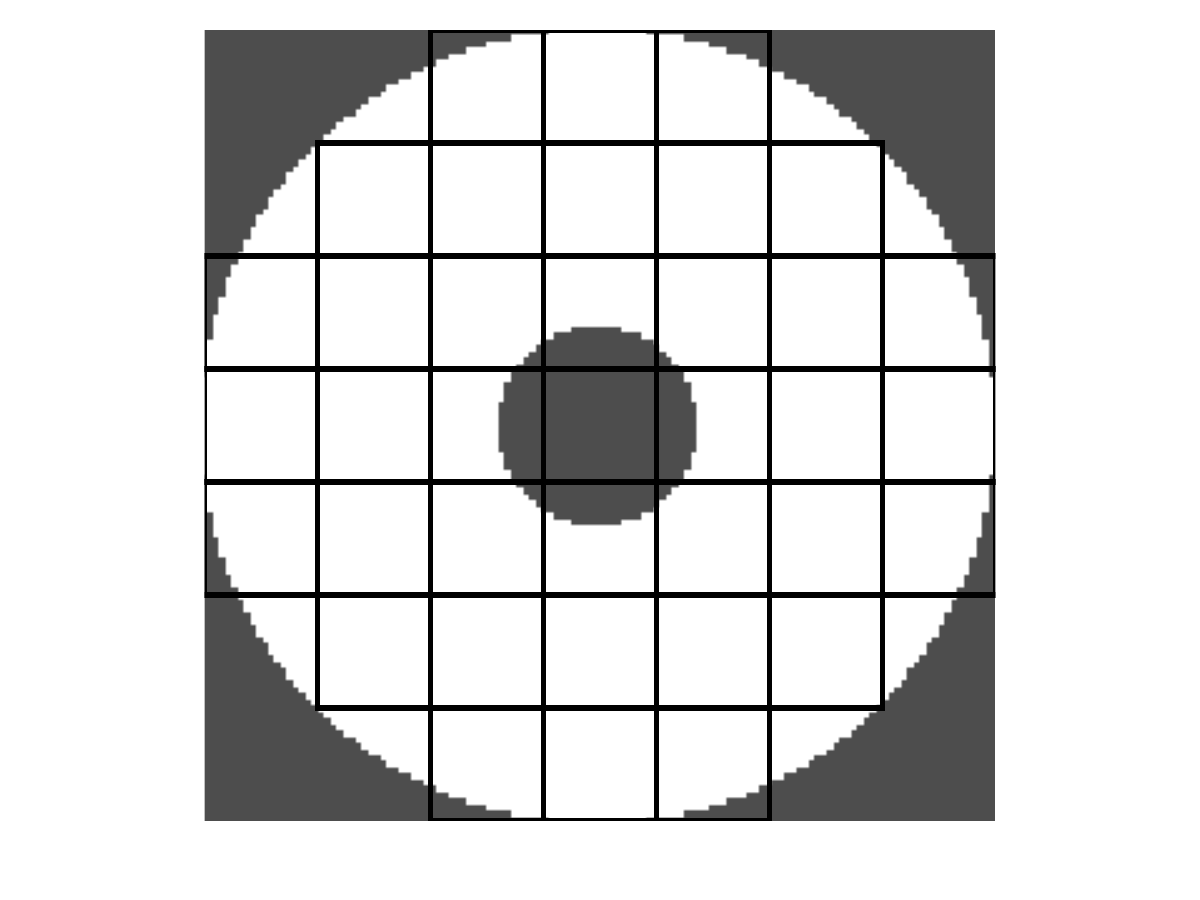}}
\caption{\label{fig:sub_apertures}
         Calculation of Shack-Hartmann \ac{WFS} sub-apertures with a given fill factor or higher for the pupil shown in Fig.~\ref{fig:circle}(b).
         Fig.~8(a) shows the creation of masks which can be applied to \ac{WFS} data to remove measurements from sub-apertures which do not receive enough light to generate useful data.
         This can be used to filter \ac{WFS} data from simulations and \acp{RTC}, removing noisy data.
         Fig.~8(b) shows the sub-aperture grid which meet required fill factor plotted over the corresponding pupil mask shown in Fig.~\ref{fig:circle}(b)}
\end{figure}

AOtools includes functions for generating sub-aperture masks for Shack-Hartmann \acp{WFS}, based on a simple square grid, by inputting the geometry of the sub-aperture grid and the pupil mask.
It is also possible to calculate the fill-factor for a set of sub-apertures for a given pupil mask.
Example code for doing both of these are given in Fig.~\ref{fig:sub_apertures}(a) and an example set of valid sub-apertures shown on the telescope pupil shown in Fig.~\ref{fig:circle}(b) is shown in Fig.~\ref{fig:sub_apertures}(b).
As well as creating sub-aperture masks for given pupil shapes, AOtools also includes methods to convert a ``stream'' of Shack-Hartmann \ac{WFS} data into a 2-D grid which matches sub-aperture geometry.
This allows for the conversion of data which comes out of \ac{RTC} systems, such as DARC \cite{Basden2010, Basden2012a}, for analysis.

\subsection{Optical propagation}
\label{sec:propagation}
\begin{figure}
\begin{mintedsubcaptionbox}{Code snippet.}{python}
>>> import numpy
>>> from aotools import opticalpropagation
>>> wavelength = 500e-9
>>> propagation_distance = 10000.
>>> # Create complex wavefront from phase screen
>>> wavefront = numpy.exp(j*phase_screen)
>>> # Propagate wavefront over 10km
>>> propagated_screen = opticalpropagation.angularSpectrum(wavefront,
                            wavelength, pxl_scale, pxl_scale,
                            propagation_distance)
>>> # Create image of propagated wavefront
>>> scintillation_pattern = numpy.abs(propagated_screen)**2.
\end{mintedsubcaptionbox}
\subcaptionbox{Scintillation pattern.}[1.\textwidth]
    {\includegraphics{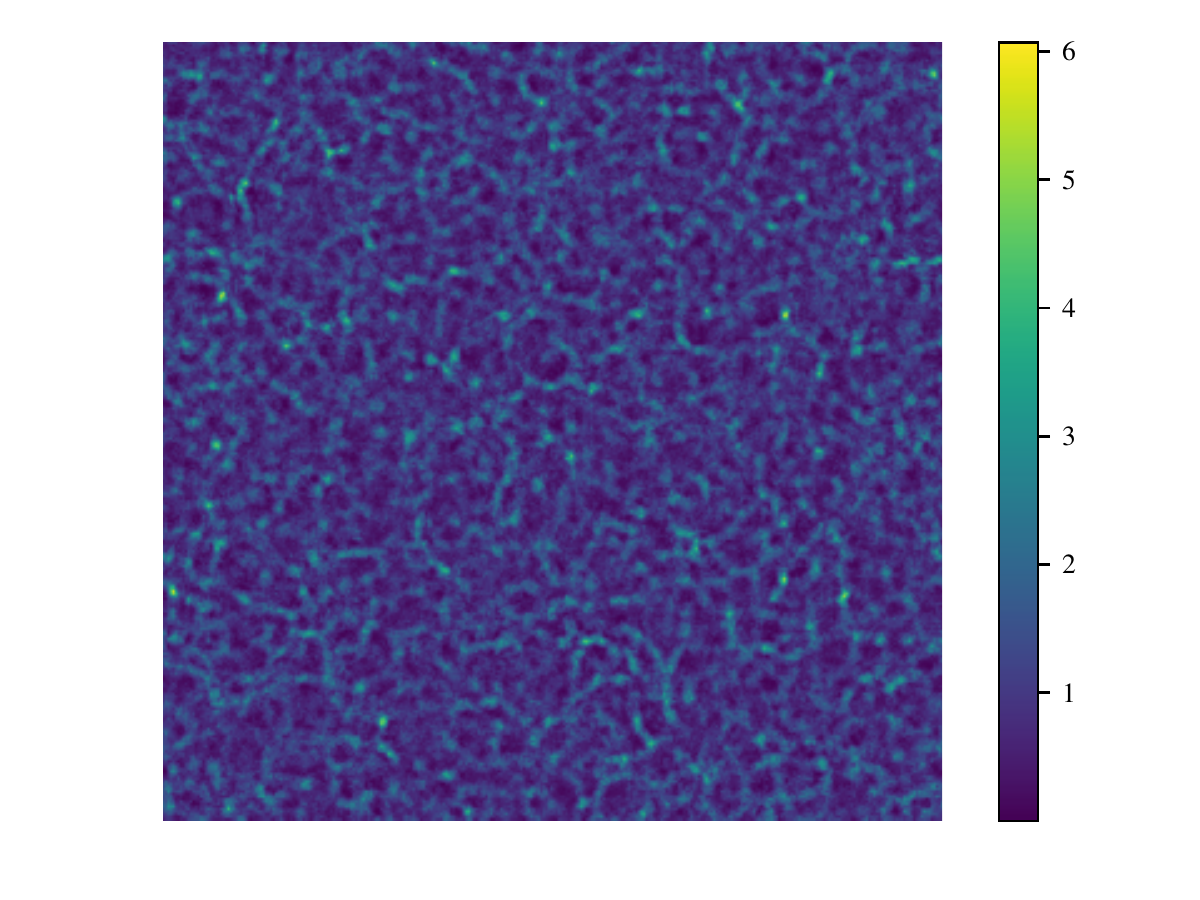}}
\caption{\label{fig:scintillation}
         Fig.~9(a) shows example code showing the propagation of the phase created shown in Fig.~\ref{fig:phase_screen}(b) for $10\si{\kilo\meter}$ and creation of the resulting scintillation image in the pupil.
         The generated scintillation pattern is shown in Fig.~9(b).}
\end{figure}

AOtools includes functions to physically propagate a wavefront through space, leading to the generation of scintillation.
Available physical propagation methods include the angular spectrum method, and one and two step Fresnel propagation.
An example code snippet of propagating a complex wavefront created from the phase screen created in Fig.~\ref{fig:phase_screen}(a) is shown in Fig.~\ref{fig:scintillation}(a).
% These functions are all physical propagation methods, leading to the generation of scintillation.
The scintillation pattern from propagating the phase screen shown in Fig.~\ref{fig:phase_screen}(b) by $10\si{\kilo\meter}$ is shown in Fig.~\ref{fig:scintillation}(b).

\begin{figure}
    \centering
    \includegraphics[width=\textwidth]{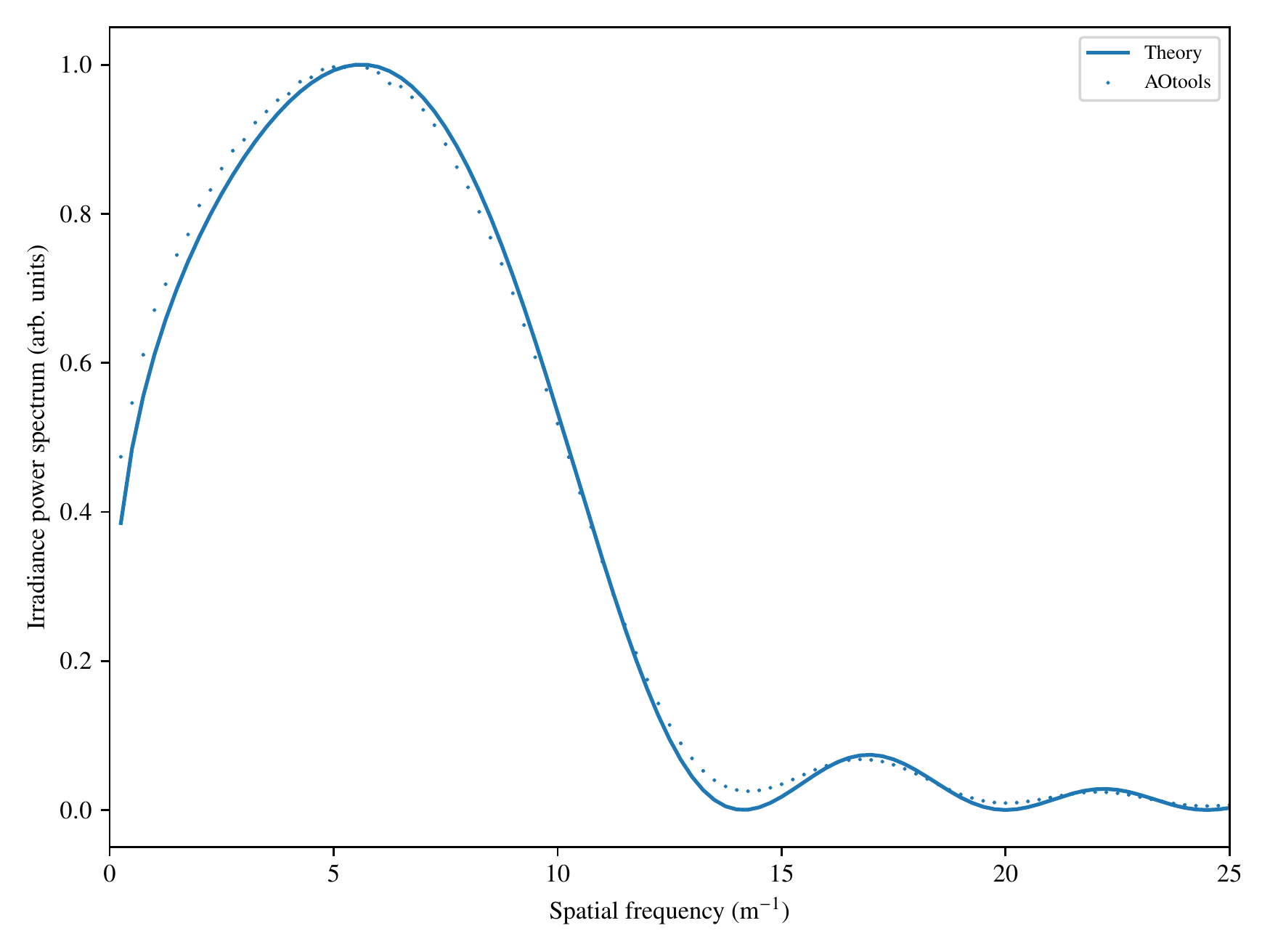}
    \caption{Scintillation irradiance power spectrum obtained by using the AOtools angular spectrum propagation to propagate 500 phase screens a distance of 10 km. Each screen has $r_0=32$ cm.}
    \label{fig:scint_powspec}
\end{figure}

In Fig.~\ref{fig:scint_powspec} we show that the scintillation irradiance power spectrum obtained through AOtools propagation is consistent with the theory outlined in \cite{Rodd1981}.

As well as physical propagation of wavefronts the optical propagation section of AOtools includes a \ac{FFT} based  function to change between the pupil plane and focal plane.
This can be used to convert a wavefront which is incident to a telescope pupil into a \ac{PSF}, as has been shown in the code snippet in Fig.~\ref{fig:psf}(a) and Fig.~\ref{fig:psf}(b).

\subsection{Astronomical unit conversions}
\label{sec:conversions}
Many standard astronomical units are not directly applicable to tasks such as \ac{AO} simulations.
For instance stellar brightnesses are typically defined through the magnitude system (see \href{http://simbad.u-strasbg.fr/simbad/}{SIMBAD}, \href{http://vizier.u-strasbg.fr/viz-bin/VizieR}{VizieR}).
However, for simulating astronomical \ac{AO} systems photon flux is required to calculate expected signal levels on \acp{WFS}.
AOtools includes conversion functions which convert stellar brightness into photon flux for a given waveband, and vice-versa.
Example code snippets are given in Fig.~\ref{code:conversions} which demonstrate converting stellar magnitude to photon flux.
\begin{figure}
\begin{minted} {python}
>>> import aotools
>>> m = 0# Use Vega as example star
>>> exp_time = 30 # 30s exposure
>>> waveband = 'V'
>>> photon_rate = aotools.magnitude_to_flux(m, waveband)
>>> # Calculate photons in an example exposure
>>> photons = aotools.photons_per_band(m, pupil_mask, pixel_scale, exp_time)
\end{minted}
\caption{\label{code:conversions}
                Example of using AOtoools to convert stellar flux from magnitudes to photon flux.
                Examples for both a raw conversion to photon rate and also for the telescope pupil shown in Fig.~\ref{fig:circle}(b) and a given exposure time.}
\end{figure}

As well as converting between astronomical units, AOtools also includes a number of parameter conversion functions which are specific to astronomical \ac{AO}.
These predominantly relate to the conversion between different atmospheric parameters, such as $Cn^{2}$, $r_{0}$, seeing, \emph{etc}.
Examples showing the conversion between these different units are shown in Fig.~\ref{code:astro}.
AOtools also includes methods to calculate integrated atmospheric parameters, such as coherence time and isoplanatic angle from atmospheric turbulence and wind speed profile.
\begin{figure}
\begin{minted} {python}
>>> import aotools
>>> r0 = 0.15
>>> cn2 = aotools.r0_to_cn2(r0)
>>> seeing = aotools.r0_to_seeing(r0)
\end{minted}
\caption{\label{code:astro}
        Example use of AOtools to convert between different typically used astronomical \ac{AO} units.
        }
\end{figure}

\section{Infrastructure and development approach}
\label{sec:dev}
AOtools is a Python library which has been developed for the AO community by the AO community.
It is hoped that this continues, and the involvement of the community will increase over time to include users of \ac{AO} beyond astronomy and microscopy.
In order to achieve this a large focus of AOtools is outside of functionality of the module and instead in infrastructure surrounding the tools.
We use the software related expertise of both the AO community and dedicated software engineers to ensure the code is well maintained, written clearly with good documentation, and well tested.
The general infrastructure goals for AOtools are to ensure that it is is well tested,  documented, and is both easy to install and easy to use.

In order to encourage input to the codebase it is hosted on \href{https://github.com/AOtools/aotools}{Github}, where it is open-source under a GNU Lesser General Public License v3.0. license \cite{LGPL}.
This license allows for any work to use AOtools, including commercial, without forcing any licence on the new software.
However, it does insist that any changes to the core AOtools package are made public under a similar license.
We utilise the infrastructure available through GitHub and the \href{http://git-scm.com}{git software} to update the codebase and to manage the release of new versions.
Changes to the codebase are submitted though ``pull requests'', where changes (whether they are bug-fixes or new features) are reviewed by core members of the AOtools team and merged into the main code base once the code has been approved.

Updating the main code base of AOtools is often followed by a release of a new version.
The versions are numbered according to the major.minor.micro format as described in \href{https://www.python.org/dev/peps/pep-0440/}{PEP440}.
Major versions are those which contain major changes and may break compatibility with previous versions, minor versions are those which contain new features or enhancements to existing features (these should not break compatibility with versions within the same Major version), and micro versions typically contain bug fixes for issues found within the existing source code.
After a new version of AOtools has been created it is then registered with the \ac{PYPI}.
This allows for AOtools be installed and updated easily by users using `pip'.
AOtools can be installed with Python using \verb+pip install aotools+.

The AOtools package makes use of continuous integration.
This is the automated running of unit tests whenever changes are made to the main code base.
As part of this testing AOtools is run on a number of different versions of Python, on a number of different operating systems in an attempt to avoid issues arising due to specific installations and environments.
We also keep track of how much of the main code base is covered by this automated testing using code coverage tools from \href{https://codecov.io/gh/AOtools/aotools}{codecov}.

AOtools has extensive documentation, which is created using the Sphinx documentation generator\cite{Brandl2010} to automatically generate documentation for every available function and module based on the docstrings in the source code.
This ensures both that the source code itself is well documented, but also generates a complete set of documentation which is then hosted in an interactive format for ease of use at \href{http://aotools.readthedocs.io}{read the docs}.
An example of the documentation is given in Fig.~\ref{fig:circle}(c) for the circle function which can be used to generate telescope pupil masks.

\section{Future development}
\label{sec:future}
AOtools is a relatively young project and in active development.
There are many updates and incremental additions to functionality ongoing and planned for the future.
Here we summarise the major updates in functionality that are currently planned: Shack-Hartmann \ac{WFS} slope analysis, and image processing.

Although there are some routines in AOtools for analysing \ac{WFS} data there are still many techniques which are not implemented.
We plan on expanding the Shack-Hartmann \ac{WFS} analysis routines available in AOtools to include measuring atmospheric parameters from the slopes \cite{Wilson2002,Butterley2006} and generation of covariance matrices from both \ac{WFS} data and atmospheric turbulence profiles.
Finally, functionality will be developed for the creation of \ac{AO} reconstructors based on covariance matrices which have been pre-calculated.

The development of image processing routines is concentrated on \acp{PSF} analysis.
AOtools will feature simple to use methods for generating \acp{PSF} using different inputs, including phase and complex amplitude in a pupil.
Building on this we will develop methods for applying a \ac{PSF} to extended images, to allow for the degradation of wide-field images by the effects of atmospheric turbulence.

The final areas development for AOtools are within the telescope pupil generation and in adding \acp{DM}.
Currently this is very simple, based on circular apertures with cut-outs for a central obscuration.
We would like to develop this to be more flexible, including the possibility for building segmented telescopes which are not completely circular.
As of yet there are no features related to the creation or modelling of \acp{DM} within AOtools.
However, this is something which could be of use to the community.

We will continue to maintain and improve much of the infrastructure of AOtools, expanding documentation to include example code demonstrating the use of the functions, much as presented in and alongside this paper.
Currently AOtools features generic \ac{AO} functions and data processing methods and functions, and some methods which apply specifically to astronomy.
We are looking to expand the functionality of AOtools to include other fields which make use of \ac{AO}, such as microscopy.
Finally, we plan on submitting AOtools to be included as an affiliated package to Astropy \cite{Robitaille2013a} to achieve better integration with the features and facilities they offer.

\section{Summary}
The AOtools package is aimed to be a core package for \ac{AO} scientists and users of data taken with \ac{AO} systems.
It is currently in use in a number of astronomical projects and institutions, however, we plan to expand beyond astronomy into the wider \ac{AO} community.
In this paper we have described the major functionality of AOtools in release 1.0, including:
\begin{itemize}
    \item Pupil functions (\S\ref{sec:pupil})
    \item Atmospheric turbulence (\S\ref{sec:turbulence})
    \item AO image processing (\S\ref{sec:image})
    \item Wavefront sensors (\S\ref{sec:wfs})
    \item Optical propagation (\S\ref{sec:propagation})
    \item Unit conversion (\S\ref{sec:conversions}).
\end{itemize}

In Sec.~\ref{sec:dev} we outlined the development approach and core ideas behind the AOtools project.
This has enabled us to begin to involve the \ac{AO} community with contributing to the project.
Finally, we outlined the future plans for the project in Sec.~\ref{sec:future}, with the major goals of increasing functionality and becoming affiliated with the Astropy project.

\section*{Funding}
\ac{STFC} (ST/L00075X/1); \ac{STFC} Studentship (ST/N50404X/1).
GOX acknowledges support from the Beware Fellowships Academia program (convention \#1610368) and the Walloon region of Belgium through the program Skywin (convention \#7751).

% \section*{Acknowledgments}
% Acknowledgments, if included, should appear at the end of the document. The section title should not be numbered.

\section*{Disclosures}
The authors declare that there are no conflicts of interest related to this article.

%%%%%%%%%% If using BibTeX:
\bibliography{OSA-template}

\end{document}